\documentclass[preprint, preprintnumbers,nofootinbib,12pt,aps]{revtex4-2}
\usepackage[english]{babel}
\usepackage{color}
\usepackage{amsmath}
\usepackage{physics}
\usepackage{soul}
\usepackage{xr-hyper}
\usepackage[hidelinks]{hyperref}
\usepackage{diagbox}
\usepackage{multirow}
\usepackage{amsfonts}
\usepackage{amssymb}
\usepackage{latexsym}
\usepackage{graphicx}
\usepackage{adjustbox}
\usepackage{float}
\usepackage{xr}
\usepackage{caption}
\usepackage{enumitem}
\usepackage{subfigure}
\usepackage{seqsplit}
\usepackage{ytableau}
\usepackage{booktabs}

\usepackage{mathtools}
\usepackage[nodisplayskipstretch]{setspace} 

\newcommand{\pa}{\partial}

\newcommand{\veps}{\varepsilon}

\newcommand{\la}{\lambda}

\newcommand{\rar}{\rightarrow}

\newsavebox\CBox
\def\textBF#1{\sbox\CBox{#1}\resizebox{\wd\CBox}{\ht\CBox}{\textbf{#1}}}

\begin{document}
\title{Two- and Three-Particle Complexes with Logarithmic Interaction: Compact wave functions for Two-Dimensional Excitons and Trions}
 \author{J.C.~del~Valle$^{1,}$\footnote{Corresponding author.\\ E-mail address: juan.delvalle@ug.edu.pl  }, J. A. Segura~Landa$^2$, and D. J.~Nader$^{3}$}
\address{
  $^1$Institute of Mathematics,  University of Gda\'nsk, ul. Wit Stwosz 57, 
  80-308 Gda\'nsk, Poland\\
  $^2$Facultad de F\'isica, Universidad Veracruzana, A. Postal 70-543 C. P. 91090, Xalapa, Veracruz, Mexico\\
  $^3$Department of Chemistry, Brown University, Providence, Rhode Island 02912, United States}
	\begin{abstract}
Assuming a logarithmic interaction between constituent particles, compact and locally accurate wave functions that describe bound states of the two-particle neutral and three-particle charged complexes  in two dimensions  are designed.   Prime examples of these complexes are excitons and trions that appear in monolayers of Transition-Metal DichalCogenides (TMDCs).
In the case of excitons, these wave functions led to 5-6 correct decimal digits in the energy and the diamagnetic shifts.  In addition, it is demonstrated that they can be used as zero-order approximations to study magnetoexcitons via perturbation theory in powers of the magnetic field strength. 
For the trion, making a comparison with experimental data for concrete TMDCs, we established that the logarithmic potential leads to binding energies $\lesssim30$\% greater than experimental ones. Finally, the structure of the wave function at small \textcolor{black}{and large} distances was  established for excitons whose carriers interact via the Rytova-Keldysh potential.  
\end{abstract}
\maketitle
\newpage
\section*{Introduction}
 It was long-time ago established \cite{Rytova,Keldysh} that a logarithmic interaction  between free electrons \textcolor{black}{$(e^{-})$} and holes $(h)$  may occur in  two-dimensional semiconductors as a limiting case of the Rytova-Keldysh potential\footnote{\textcolor{black}{See Section III for a detailed discussion based on excitons.}}. 
 Let us consider a thin  film/layer with a dielectric surrounding made of  dielectric substrates. 
 If the thickness is sufficiently small compared with the exciton Bohr radius, a logarithmic interaction  between carriers confined to the film  emerges as a result of the polarization of atomic orbitals \cite{Aleiner}. 
 Monolayers of Transition-Metal 
DichalCogenides (TMDCs) are relevant examples of this kind of film-substrate configurations.  Such
monolayers can be considered as  two-dimensional since their thickness is a few Angstroms \cite{Yang}. The low dimensionality (planar) and  dielectric screening in such materials result in a strong electrostatic interaction. It allows the existence of stable bound states for  complexes composed by electrons and holes, such as excitons \textcolor{black}{($e^{-},h$)}, and  positively and negatively charged trions: \textcolor{black}{$(e^{-},h,h)$} and  \textcolor{black}{$(e^{-},e^{-},h)$}, respectively. Monolayers of TMDCs have recently received special attention for being ideal candidates for potential applications in optoelectronics \cite{Ross,Wang}, valleytronics \cite{Liu},  enhanced photoluminescence \cite{Splendiani,Mak}, and systems with pronounced
many-body effects \cite{Ross}. In particular, the optical response of these materials is explained in terms of complexes \cite{Aleiner}. 
For a multilayer configuration of TMDCs, the pair-wise  interaction between two  carriers of charge $q_i$ and $q_j$ within the same  layer remains logarithmic\footnote{\textcolor{black}{For a detailed discussion and derivation of the interaction see  \cite{Formin}, the supplemental material of \cite{Aleiner}}.}, having the same form as for monolayers:
	\begin{equation}
		V_{ij}(\rho)\ =\ -\frac{q_iq_j}{\rho_0}\ln\left(\frac{\rho}{\rho_0}\right)\ .
		\label{potential}
\end{equation}
 Here,  $\rho$ denotes the  relative distance between the carriers, whose electric charges are $q_{i}$ and $q_{j}$, and 
 \begin{equation}
     \rho_0\ =\frac{d\, (\varepsilon_{||}-1)}{2}\ ,
 \end{equation}
 where $\varepsilon_{||}$ is the in-plane component of the dielectric permittivity tensor of the bulk layered material, and $d$ is the distance between layers.  

In the effective-mass approximation, the quantum mechanical description of  complexes is governed by the Schr\"odinger equation. In this context, the variational method has shown to be an adequate tool to study bound states of excitons and trions, see \cite{Martins,Grasselli,Zhang_2019,Semina,Molas,Pedersen} and references therein.
However, as mentioned in  \cite{Komsa},  difficulties in constructing a reasonable wave function
Ansatz in the case of  larger complexes than excitons hinders the straightforward extension of the variational consideration. To overcome this drawback, some recent advances have been made in constructing adequate wave functions for \textcolor{black}{intralayer} trions  \cite{Courtade,Semina}.  However, most of the variational functions favor the simplicity in  calculations leading to reasonable results in terms of energy, instead of a correct description of the wave function. The main goal of this work is to show  that, for two/three-particle complexes, both accurate energies and wave functions  can be simultaneously achieved by using adequate compact trial functions.

In the present study, we construct  compact  Ans\"atze (trial functions)  associated with the  intralayer exciton of multi and monolayers made of TMDCs using  the \textit{internal} structure of the exact wave functions in the logarithmic regime of interaction. For unclear reasons to the authors, this approach has not been studied so far in a variational consideration.  
We focus on the construction of locally accurate approximations of the wave functions. They are valuable  not only for finding  energies and  expectation values with high accuracy, but also for shedding some light on physical properties of the exact wave functions. 
 In fact, the search for compact wave functions describing few-particle charged systems is an active field of research \cite{Bressanini_2008}. For example, they are widely used in atomic physics due to their usefulness to compute efficiently scattering cross sections \cite{PhysRevLett.128.053001}
 and matrix elements  of  singular operators \cite{Yerokhin2021}. In quasi-two-dimensional materials, the slow convergence of CI (configuration interaction) functions \cite{Planelles,Quintela,plane2021} has motivated the search for simple yet accurate wave functions for the description of hole-electron interaction.

As we  show, our approximate  solutions (taken as zero order approximation)  lead to  a convergent perturbation series to the exact solution.
Two concrete physically relevant examples of application are discussed:
(i)   magneto-excitons in a weak field regime; and most importantly (ii) the
construction of  three-particle  wave functions.




 The present work is organized as follows. In Section \ref{SectionI}, we discuss the construction of  compact parameter-dependent exciton wave functions. Concrete variational calculations for the spectrum of the first low-lying states are presented. We investigate the accuracy of the energy estimates using the non-linearization procedure \cite{Turbiner1984} and an alternative variational trial function.  Then, we study the effect of a weak uniform magnetic field to the energy spectrum using perturbation theory taking our compact functions as zero order approximation. In this line, we discuss the (re)summation of perturbation series using Pad\'e approximants. 
 In Section \ref{SectionII},
 taking as building block the trial function constructed for the exciton, we propose a trial function for the ground state of the trion. We study the binding energy of the complex for concrete  TMDCs and compare it with experimental results.  A simple formula  for the trion energy is provided. In Section \ref{SectionIII}, we discuss extensions of our consideration to the Rytova-Keldysh potential.  Finally, we summarized our results in Conclusions.

 \section{Two-Particle System: Excitons}
 \label{SectionI}
	Consider the neutral system made
 of two charged particles (hole and electron) interacting through  potential (\ref{potential}). After separating the motion of the  center of mass and using  polar coordinates, we arrive at the familiar two-dimensional radial Schr\"odinger equation for the relative motion,
	\begin{equation}
		-\frac{\hbar^2}{2\mu}\left(\pa_\rho^2\psi\ +\ \frac{1}{\rho}\pa_\rho\psi\right)\ +\ \left(\frac{\hbar^2m^2}{2\,\mu\rho^2}\ +\ \frac{e^2}{\rho_0}\ln\left(\frac{\rho}{\rho_0}\right)\right)\psi\ =\ E\,\psi \ ,\qquad 
		\rho\in[0,\infty)\ ,
		\label{Schroedinger}
	\end{equation}
	where $\mu$ is the reduced mass\footnote{By definition $\mu\ =\ \dfrac{m_e\,m_h}{m_e+m_h}$, where $m_e$ and $m_h$ are the electron and hole effective masses, respectively.},  $e$ denotes the charge of the hole, and $\hbar$ is the reduced Planck constant. Any  energy and wave function can be labeled by ($n_\rho,m$), but for simplicity we drop such labels for now. The radial quantum number takes the values $n_\rho=0,1,...$, meanwhile the magnetic quantum number $m=0,\pm1,...$ .  Using the transformation \textcolor{black}{via the dimensionless coordinate}
	\begin{equation}
		\rho\rar\left(\frac{\hbar^2\rho_0}{\mu\,e^2}\right)^{-\frac{1}{2}}\rho\ ,
		\label{transformation}
	\end{equation}
	we remove the explicit presence of the constants $\hbar$, $\mu$, $\rho_0$,  and $e$ from the Schr\"odinger equation \textcolor{black}{which is transformed into its dimensionless form}
	\begin{equation}
		-\frac{1}{2}\left(\pa_\rho^2\psi\ +\ \frac{1}{\rho}\pa_\rho\psi\right)\ +\ \left(\frac{m^2}{2\,\rho^2}\ +\ \ln\left(\rho\right)\right)\psi\ =\ \varepsilon\,\psi \ .
		\label{key}
	\end{equation}
	 In this equation, $\varepsilon$ plays the role of \textcolor{black}{dimensionless} energy, and it is related to $E$ through
	\begin{equation}
		E\ =\ \frac{e^2}{\rho_0}\,\varepsilon \ -\ \frac{e^2}{2\rho_0}\,\ln\left(\frac{\mu\,e^2\rho_0}{\hbar^2}\right)\ .
		\label{energyE}
	\end{equation}
	The second term in (\ref{energyE}) only provides the reference point to measure energies, and it has no relevant role.  From (\ref{energyE}), it is clear that the energy difference between two arbitrary states does not depend on the reduced  mass $\mu$. 	Since we are interested in bound states, we impose boundary conditions on (\ref{key}) such that the normalizability requirement
	\begin{equation}
		\int_{0}^{\infty}|\psi(\rho)|^2\rho\,d\rho\ <\ \infty
	\end{equation}
	is fulfilled. Under these considerations, equation (\ref{key}) does not admit an exact solution: energies and wave functions can only be  found in approximate form. Interestingly, the same spectral problem defined by equation (\ref{key}) can appear in another context. Indeed, the effect of a wiggly cosmic string for both mass-less and massive particle propagation along the string axis is governed by (\ref{key}), see \cite{Azevedo}.
 
	\subsection{ Ground State}
	Relevant information concerning the structure of the wave function can be revealed using asymptotic analysis.  It is convenient to adopt the  \textit{exponential representation} of the wave function, namely
 	\begin{equation}
		\psi_{n_\rho,m}(\rho)\ =\ \rho^{|m|}P_{n_\rho,m}(\rho^2)\exp(\,-\Phi_{n_\rho,m}(\rho)\,)\ 
  \label{representation}.
	\end{equation}
 The unknown function $\Phi_{n_\rho,m}(\rho)$  is called  \textit{phase}, while  $P_{n_\rho}$ is a polynomial of degree $n_\rho$. Explicitly,
 	\begin{equation}
		P_{n_\rho}(\rho^2)\ =\ \prod_{i=1}^{n_\rho}(\rho^2-\rho_i^2)\ ,\qquad P_0(\rho)\ \equiv\ 1\ .
  \label{prefactor}
	\end{equation}
 For a given state, this polynomial is determined by the position of the nodes $\rho_i$, $i=1,...,n_\rho$. Thus,  representation (\ref{representation}) is unambiguous \cite{Turbiner1984}. The asymptotic series of the ground state phase, $\Phi_{0,0}(\rho)$, shares properties with those for excited states. 
 Hence, we focus on the quantum numbers ($n_\rho=0,m=0)$ from now on. In this case, (\ref{representation}) takes the form
	\begin{equation}
		\psi_{0,0}(\rho)\ =\ \exp(\,-\Phi_{0,0}(\rho)\,)\  .
	\end{equation}
	We  construct the asymptotic series  around  two relevant points of the domain:  $\rho=0$ and $\rho=\infty$; small and large relative distances, respectively. At  $\rho=0$, it can be demonstrated that the asymptotic series of the phase has the following structure
	\begin{equation}
		\Phi_{0,0}(\rho)\ =\   \sum_{i=1}^{\infty}\,\sum_{j=0}^ia_{ij}\rho^{2i}\ln^{j}(\rho)\ ,\qquad \rho\ \rar\ 0\ ,
		\label{series1}
	\end{equation}
	where $a_{ij}$ are coefficients with dependence on $\varepsilon$. 
	As a consequence, the  wave function has a similar asymptotic series,
	\begin{equation}
		\psi_{0,0}(\rho)=  \sum_{i=0}^{\infty}\,\sum_{j=0}^ib_{ij}\rho^{2i}\ln^{j}(\rho)\ ,\qquad \rho\ \rar\ 0\ ,    
		\label{series2}
	\end{equation}
	where $b_{ij}$ are $\varepsilon$-dependent coefficients. On the other hand, the  
	first terms of the  asymptotic series of the  phase at $\rho=\infty$ are
	\begin{equation}
		\Phi_{0,0}(\rho)\ =\ \rho\left(\sqrt{2\ln\rho\ } \ +\  \frac{1-\varepsilon}{\sqrt{2\ln\rho}}\ +\ \mathcal{O}\left((\ln\rho)^{-3/2}\right)\right) +\ \frac{1}{2}\ln\rho\ +\ ...\ ,\qquad \rho\rar\infty
		\label{series3}
	\end{equation}
	Note that the dominant term, $\mathcal{O}(\rho\sqrt{\ln \rho})$, does not depend on the energy. Therefore, the same leading term is expected for any bound state. Contrary to the series at  $\rho=0$, see (\ref{series1}), we were unable to find the general structure of the phase at $\rho =\infty$. However, for the particular purpose of this work, this piece of information is not required. 
At $\rho\rar0$, the phase of any state has the same structure of series (\ref{series1}).
On the other hand, the wave function of any  state has the following asymptotic expansion,
 	\begin{equation}
 	\psi_{n_\rho,m}(\rho)=  \rho^{|m|}\sum_{i=0}^{\infty}\,\sum_{j=0}^ib_{ij}\rho^{2i}\ln^{j}(\rho)\ ,\qquad \rho\ \rar\ 0\ ,    
 	\label{seriesexcited}
 \end{equation}
 with coefficients $b_{ij}$ depending on $\varepsilon_{n_\rho,m}$, and $|m|$.  
 Using series (\ref{seriesexcited}), the real solutions of  the equation $\psi_{n_\rho,m}(L)=0$, with $L>0$ sufficiently large, define low-accuracy approximations to the low-lying  energies and wave functions.  
	
	\subsection{Compact Trial Function}

	Based on series (\ref{series1}), (\ref{series2}), and (\ref{series3}), we constructed a parameter-dependent approximation for the wave function of an arbitrary state $(n_\rho,m)$. We followed the prescription described in \cite{Turbiner_2021}, where it was applied successfully to the quartic anharmonic oscillator. Such prescription establishes the following: the approximate phase is the result of  matching the series  (\ref{series1}) and (\ref{series3}) in a minimal way, reproducing as many dominant growing terms in  (\ref{series3}) as possible. This procedure is almost unambiguous, and it leads to
	\begin{equation}
		\Phi_{n_\rho,m}^{(approx)}(\rho)\ = \ 	\dfrac{A\ +\ (1-2B)\,\rho^2\ln\rho\ +\ (C+B\rho^2)\ln(1+D\rho^2)}{\sqrt{F^2\, (1+D\rho^2)+\dfrac{1}{2}\rho^2\ln \rho}}\ -\ \frac{A}{F}\ ,
		\label{phase}
	\end{equation}
	where $\{A,B, C,D ,F\}$  are five  $(n_\rho,m)$-dependent  \textcolor{black}{dimensionless} parameters.  By construction, the phase (\ref{phase}) reproduces functionally (same structure, but different coefficients) all the terms in the expansion at small distances (\ref{series1}), but only the leading one in the expansion at large distances (\ref{series3}).  The approximate trial wave function of an arbitrary state ($n_\rho,m$) is given ultimately by	
	\begin{equation}
		\small
		\psi_{n_\rho,m}^{(approx)}(\rho)\ =\ \rho^{|m|}P_{n_{\rho}}(\rho^2)\times \exp\left(-\dfrac{A\ +\ (1-2B)\,\rho^2\ln\rho\ +\ (C+B\rho^2)\ln(1+D\rho^2)}{\sqrt{F^2\, (1+D\rho^2)+\dfrac{1}{2}\rho^2\ln \rho}}\ +\ \frac{A}{F}\right) \ ,
		\label{trial}
	\end{equation}
	where the polynomial $P_{n_{\rho}}(\rho^2)$ is of the form (\ref{prefactor}), and it carries the information about the nodes. To fix the value of free parameters, we use the variational method imposing the  orthogonalization constraints
	\begin{equation}
		\langle \psi_{n_{\rho,m}}^{(approx)}|\psi_{n'_\rho,m}^{(approx)} \rangle \ =\ 0,\qquad n'_\rho= 0,1,...,n_{\rho}\ .
		\label{constraints}
		\end{equation}
	These constraints define the position of the nodes.  Thus, after fixing $m$, we move sequentially from the ground state $(n_\rho=0)$ to a higher excited state $n_\rho>0$. Under these constraints, we calculate the parameter-dependent expectation value of the Hamiltonian associated with (\ref{key}), usually called  \textit{variational energy} and denoted by $\varepsilon^{(var)}_{n_\rho,m}$. Then,  using an optimization procedure, we can find the  configuration of parameters  that minimize the variational energy. Only for  states with quantum numbers $(n_\rho=0,m)$, the variational principle guarantees that  the variational energy  is an upperbound of the exact energy.  To ensure the square-integrability of (\ref{trial}), the constraint $D>0$ is imposed. 
	
	Concrete numerical calculations were carried out  for the first low-lying states with $n_\rho\leq 6$ and $|m|\leq 4$.  As a result, the variational energy was found with relative accuracy $\sim 10^{-6}$ or less.  In Table \ref{groundstate}, we present the optimal parameters and the variational energies  for the first seven  $S$-states.  Meanwhile, nodes are obtained with 5 exact significant digits. It was confirmed by using the non-linearization procedure and making alternative accurate variational calculations, see below.   The optimized variational energies reach and sometimes overcome the best results found in the literature \cite{Asturias,Eveker,Gesztesy,Aleiner}.

		\begin{table}[h]
		\caption{\textcolor{black}{Dimensionless} optimal parameters and variational energies $\varepsilon_{n_\rho,0}^{(var)}$ of the first six  $S$-states. Diamagnetic shifts are also presented.  Displayed numbers are rounded.}
		{\setlength{\tabcolsep}{0.4cm}
			\begin{tabular}{|c|ccccc|cc|}
				\hline
				\hline
				\rule{0pt}{3ex}$n_\rho$& $A$ & $B$ & $C$ &$D$ & $F$& $\varepsilon_{n_\rho,0}^{(var)}$&$\frac{1}{8}\expval{\rho^2}$ \\[4pt]
				
				\hline
				
				\rule{0pt}{4ex}0
				&-0.2259&0.7684&0.1879&1.9102&0.9561&0.179\,935&0.1363\\[5pt]
				1&-2.1430&0.9613&-0.0655&0.7327&1.5413&1.314\,677&1.0363\\[5pt]
				2&-3.5415&1.0256&-0.3938&0.6052&1.8695&1.830\,608&2.8459\\[5pt]
				3&-4.6212&1.0659&-0.6910&0.5576&2.1177&2.168\,874&5.5630\\[5pt]
				4&-5.4135&1.0935&-0.9265&0.5393&2.3216&2.421\,054&9.1873\\[5pt] 
				5&-6.1751&1.1193&-1.1668&0.5214&2.4880&2.622\,221&13.7150\\[5pt] 
				6&-6.9278&1.1438&-1.4163&0.5044&2.6286&2.789\,590&19.1070\\[8pt] 
				\hline \hline
		\end{tabular}}
		\label{groundstate}
	\end{table}
	  By construction, our wave functions $\psi_{n_{\rho,m}}^{(approx)}$ are locally accurate, mainly  due to the correct asymptotic behavior of the wave function at small and large distances. To check this,  we can estimate the local accuracy of $\psi_{n_\rho,m}^{(approx)}(\rho)$ using the non-linearization procedure, via the perturbation series \cite{Turbiner1984}:
	  \begin{equation}
	  	\psi_{n_\rho,m}^{(exact)}\ =\ \psi_{n_\rho,m}^{(approx)}(1 - \phi_1  - \phi_2...)\ .
	  	\label{PTphase}
	  	\end{equation}
	  Furthermore, the non-linearization procedure dictates that 
	  \begin{equation}
	  	\varepsilon_{n_\rho,m}^{exact}\ =\ \varepsilon_{n_\rho,m}^{(var)}\ +\ \varepsilon_2+\ \varepsilon_3\ +\ ...\ ,
	  	\label{PT}
	  	\end{equation}
	  if $\psi_{n_\rho,m}^{(approx)}$ is chosen according to the above mentioned prescription. Numerical calculations for $S$-sates established that corrections $\phi_n$ in (\ref{PTphase}) decrease as $n$ grows, specifically $|\phi_{n+1}/\phi_n|\lesssim10^{-1}$ for all $\rho$. Therefore, the local accuracy of wave functions is guaranteed.  
	  In turn,  $|\varepsilon_{n+1}/\varepsilon_{n}|\lesssim10^{-2}$, which indicates a fast rate of convergence of (\ref{PT}). In particular, the value of $\varepsilon_2$ suggests that our variational calculations lead to energies with 5 - 6 exact decimal digits. 
	 For the states considered, corrections in (\ref{PTphase}) and (\ref{PT}) were calculated using the Mathematica codes described in \cite{CodeMathematica}.
	  In Table \ref{Corrections}, we present explicit values of the first 11 corrections  $\varepsilon_{n}$ in (\ref{PT})  for the ground state (0,0). \textcolor{black}{For this  state, it is enough to consider the first three terms in (\ref{PT}) to reach the accuracy  provided by the finite-element calculations, see \cite{Mostaani}.} 
\begin{table}[h]
	\caption{Logarithmic Potential, the Ground State $(0,0)$: First 10 sums of energy corrections. Digits in bold are invariant with respect
		to the next order correction. The value of $\varepsilon_{exact}$ was calculated with the function (\ref{hydrogen}). }
	\resizebox{\textwidth}{!}{
	{\setlength{\tabcolsep}{0.30cm}
		\begin{tabular}{|cc|cc|}
			\hline\hline
			\rule{-1pt}{04ex}
			Approximation	& Value& Correction&Value  \\[5pt]
			\hline
			\rule{-2pt}{4ex} 
			$\veps_0+\veps_1$ &\textBF{0.179\,935\,4}52\,016\,501\,408\,229\,152&-$\varepsilon_2$&5.1787914339321731$\times10^{-8}$
			\\[5pt]
			$\veps_0+\veps_1+\veps_2$ &\textBF{0.179\,935\,400}\,228\,587\,068\,907\,420&$\varepsilon_3$&9.8170705093752$\times10^{-11}$
			\\[5pt]
			$\veps_0+\ldots+\veps_3$ &\textBF{0.179\,935\,400\,32}6\,757\,774\,001\,172&-$\varepsilon_4$&8.61541999679$\times10^{-13}$
			\\[5pt]
			$\veps_0+\ldots+\veps_4$ &\textBF{0.179\,935\,400\,325}\,896\,232\,001\,493&$\varepsilon_5$&9.200491318$\times10^{-15}$
			\\[5pt]
			$\veps_0+\ldots+\veps_5$&\textBF{0.179\,935\,400\,325\,905}\,432\,492\,811&-$\varepsilon_6$&1.16392216$\times10^{-16}$
			\\[5pt]
			$\veps_0+\ldots+\veps_6$ &\textBF{0.179\,935\,400\,325\,905\,31}6\,100\,595&$\varepsilon_7$&1.681331$\times10^{-18}$
			\\[5pt]
			$\veps_0+\ldots+\veps_7$ & \textBF{0.179\,935\,400\,325\,905\,317\,7}81\,927&-$\varepsilon_8$&2.7053$\times10^{-20}$
			\\[5pt]
			$\veps_0+\ldots+\veps_8$&\textBF{0.179\,935\,400\,325\,905\,317\,75}4\,874&$\varepsilon_9$&4.76$\times10^{-22}$
			\\[5pt]
			$\veps_0+\ldots+\veps_9$& \textBF{0.179\,935\,400\,325\,905\,317\,755\,3}50&-$\varepsilon_{10}$&9.0$\times10^{-24}$
			\\[5pt]
			$\veps_0+\ldots+\veps_{10}$&\textBF{0.179\,935\,400\,325\,905\,317\,755\,341}&$\varepsilon_{11}$&1.8$\times10^{-25}$
			\\[5pt]
			\hline
			\hline
			\rule{-2pt}{4ex} 
			$\veps_{exact}$& 0.179\,935\,400\,325\,905\,317\,755\,341\\[5pt]
			\cline{1-2}
	\end{tabular}}}
	\label{Corrections}
\end{table}

	\begin{table}[h]
		\caption{Energies ($\varepsilon$) of the first  Low-Lying States with quantum numbers $n_\rho\leq4$ and $|m|\leq 5$, see (\ref{energyE}). Results were obtained with $N=12$, see (\ref{hydrogen}). Underlined digits correspond to those digits reproduced by  the variational energies.   All printed digits are exact: confirmed with $N=13$.}
			{\setlength{\tabcolsep}{0.30cm}
				\begin{tabular}{|c|cc|}
					\hline
					\hline
					\rule{0pt}{4ex}\diagbox{$n_\rho$}{$|m|$} & 0 &  1 \\[4pt]
					
					\hline
					\rule{0pt}{4ex}0&
					\underline{0.179\,935\,4}00\,325\,905\,317\,755\,341&\underline{1.039\,612}\,607\,367\,968\,583\,608\,037 \\[5pt]
					1&\underline{1.314\,677}\,846\,047\,317\,635\,438\,844&\underline{1.662\,90}1\,190\,508\,306\,406\,113\,371\\[5pt]
					2&\underline{1.830\,608}\,839\,744\,414\,785\,298\,073&\underline{2.047\,765}\,063\,110\,404\,237\,580\,088 \\[5pt]
					3&\underline{2.168\,874}\,146\,054\,584\,411\,366\,434& \underline{2.326\,094}\,048\,304\,208\,876\,062\,166\\[5pt]
					4&\underline{2.421\,054}\,965\,033\,637\,757\,825\,116&\underline{2.544\,033}\,274\,577\,971\,448\,108\,208\\[5pt]
					\hline \hline
					\rule{0pt}{4ex}\diagbox{$n_\rho$}{$|m|$} & 2 &  3 \\[4pt]
					\hline
					\rule{0pt}{4ex}0&\underline{1.497\,798}\,460\,867\,032\,070\,612\,310&\underline{1.811\,273}\,253\,112\,008\,598\,564\,854 \\[5pt]
					1&\underline{1.929\,287}\,879\,273\,176\,751\,640\,227&\underline{2.141\,542}\,186\,466\,426\,635\,213\,589\\[5pt]
					2&\underline{2.233\,478}\,680\,963\,778\,182\,582\,480&\underline{2.392\,481}\,446\,049\,754\,044\,603\,719 \\[5pt]
					3&\underline{2.467\,89}6\,772\,583\,752\,523\,081\,303&\underline{2.594\,39}2\,795\,360\,084\,586\,946\,462\\[5pt]
					4&\underline{2.658\,38}9\,492\,811\,470\,794\,957\,093&\underline{2.763\,108}\,473\,682\,027\,259\,596\,914\\[5pt]
					\hline \hline
					\rule{0pt}{4ex}\diagbox{$n_\rho$}{$|m|$} & 4 &  5 \\[4pt]
					\hline
					\rule{0pt}{4ex}0&\underline{2.049\,706}\,164\,599\,668\,581\,995\,749&\underline{2.242\,142}\,115\,820\,400\,875\,545\,543 \\[5pt]
					1&\underline{2.317\,307}\,924\,417\,713\,851\,799\,756&\underline{2.467\,09}7\,474\,123\,876\,210\,404\,164\\[5pt]
					2&\underline{2.530\,69}6\,526\,629\,787\,083\,928\,815&\underline{2.652\,64}5\,890\,317\,996\,675\,838\,214 \\[5pt]
					3&\underline{2.707\,805}\,246\,985\,471\,371\,167\,548&\underline{2.810\,28}2\,701\,535\,865\,116\,991\,903\\[5pt]
					4&\underline{2.859\,00}9\,046\,649\,596\,541\,890\,608&\underline{2.947\,16}7\,432\,155\,959\,952\,276\,143\\[5pt]
					\hline \hline
		\end{tabular}}
		\label{energies hydrogen}
	\end{table}



		\subsection{Alternative  Trial Function}
	\label{Accurate_Spectrum}
	The orthogonalization procedure used to determine the position of the nodes is  accurate but impractical for highly excited states. For a given $m$, the $n_\rho$ excited state requires $n_\rho-1$ constraints in order to fulfill (\ref{constraints}).   To overcome this drawback, an alternative and efficient procedure is discussed in this Section. This approach only requires the knowledge of the ground state functions $(0,m)$.  Using the wave functions $\psi_{0,m}^{(approx)}(\rho)$ with optimal parameters, we construct the expansion 
	\begin{equation}
		\psi_{n_\rho,m}(\rho)\ \approx\ \psi^{(approx)}_{0,m}(\rho)\times \sum_{i=0}^{N}\sum_{j=0}^{i}c^{(n_\rho,m)}_{i,j}\,\rho^{2i}\,\ln^j(\rho)
		\label{hydrogen}
	\end{equation}
to describe any state.
    The factor $\psi^{(approx)}_{0,m}$ guarantees the correct   the  asymptotic dominant behavior at large $\rho$ in our trial function (\ref{hydrogen}).  Based on the series at small $\rho$, an additional factor is introduced in the form of a partial sum of (\ref{series2}) in order to improve the small-$\rho$ behavior. 
In this representation, any  wave function (\ref{hydrogen}) contains  $(N+1)(N+2)/2$ terms. Using the linear variational principle, it is known that the energies and coefficients  $c_{i,j}^{(m)}$ are determined by the secular equations (Rayleigh-Ritz method). Thus, this alternative procedure is a two-step variational consideration. To solve the secular equations,  we use the  L\"owdin orthogonalization procedure \cite{Lowdin} since the set of functions  $\{\rho^{2i}\ln^j\rho\}$ is not orthogonal. 
 In Table \ref{energies hydrogen}, we present the energies of the low-lying states with quantum numbers $n_\rho\leq4$ and $|m|\leq5$ using $N=12$. For all states considered, numerical results indicate   that the rate of convergence is about 3-4 correct digits with an increment of $N$ to $N+1$. In fact, using $N=12$, we established 24 exact decimal digits for energies of all states considered. Comparing with our variational results, we established that our locally accurate  approximations (\ref{trial}) lead to energies with 5-6 exact decimal digits, which is in agreement with our calculations via the non-linearization procedure.

	\subsection{Magnetoexcitons: S-States}
\begin{figure}
		\includegraphics[width=0.65\textwidth]{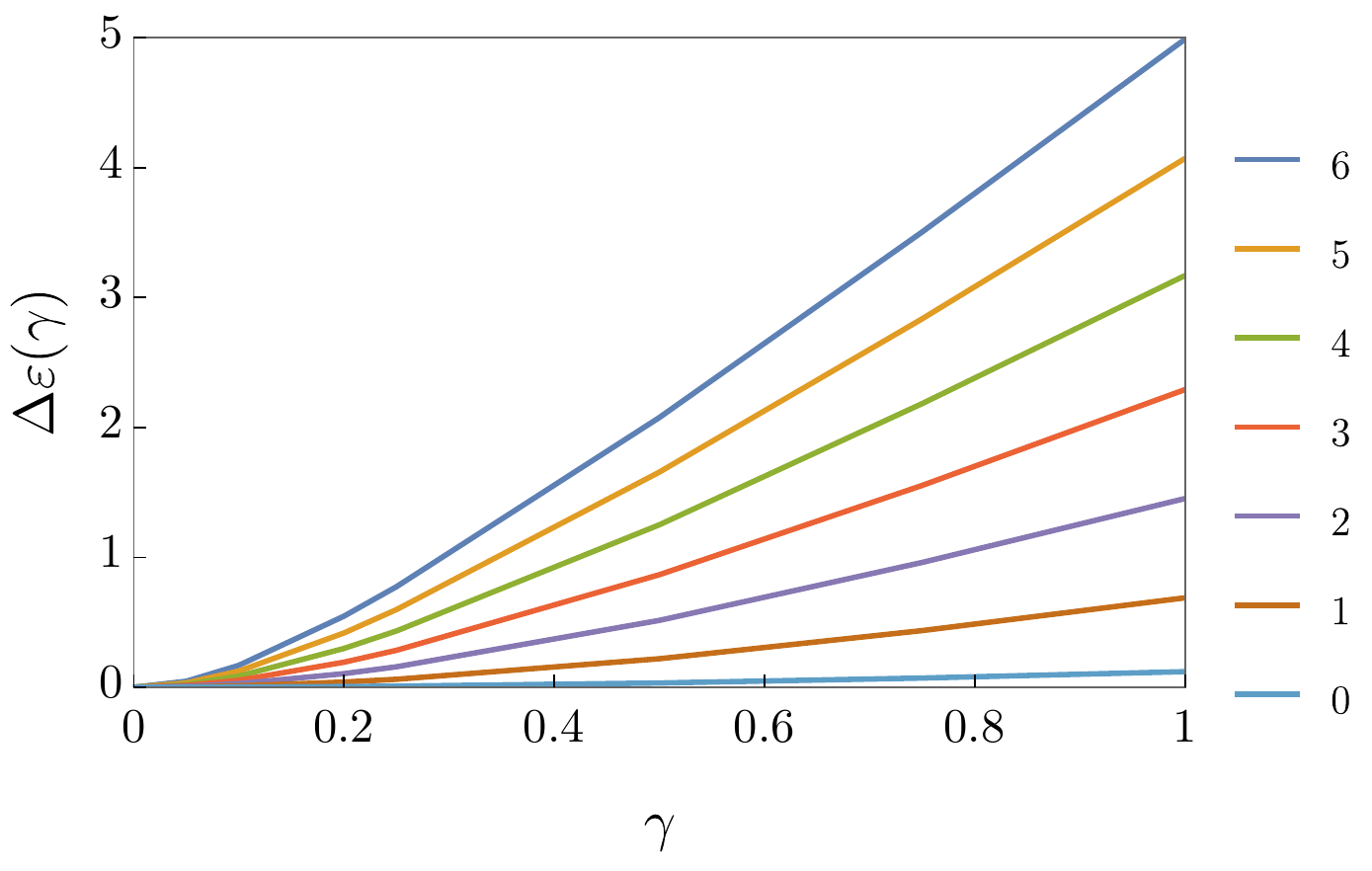}
		\caption {Plot of $\Delta \varepsilon(\gamma)$ of magnetoexcitons with quantum numbers $0\leq n_\rho\leq6$ for $\gamma\in[0,1]$. Plots constructed via Pad\'e approximants $P_{5}^6(\gamma^2)$ of the perturbation series of each state.}
		\label{Fig:MagnetoExcitons}
	\end{figure}

	Consider an exciton   subjected to a time-independent magnetic field $\boldsymbol{B} = B \hat{\boldsymbol{z}}$.\footnote{Therefore, the magnetic field is transversal to the thin film.}  As long as the momentum of the center of mass is zero,
 it can be shown \cite{Escobar2D,Herold,Magnetoexcitons} that the Schr\"odinger equation\footnote{In the symmetric gauge.} for the relative motion  describing $S$-states is
	\begin{equation}
		-\frac{1}{2}\left(\pa_\rho^2\psi\ +\ \frac{1}{\rho}\pa_\rho\psi\right)\ +\ \left(\ln\left(\rho\right)\ +\ \frac{\gamma^2}{8}\rho^2\right)\psi\ =\ \varepsilon\,\psi \ .
		\label{EinB}
	\end{equation}
 where 
 \begin{equation}
   \gamma\ =\ \frac{B}{B_0}\ ,\qquad B_0\ =\ \frac{c\,e^3\, \mu^2}{\hbar^3}\ .  
 \end{equation}
Equation (\ref{EinB}) is written in variable (\ref{transformation}).  Note that $B_0$   is the exciton unit of magnetic field while $\varepsilon$ is defined through (\ref{energyE}). For weak magnetic fields, perturbation theory for $\varepsilon({\gamma})$ can be constructed in the form
\begin{equation}
\varepsilon(\gamma)\ =\ \varepsilon_0\ +\ \sum_{n=1}^{\infty}\varepsilon_n\gamma^{2n}\ .
\label{Perturbation}
\end{equation}
The first order correction ($\varepsilon_1$), given  by
\begin{equation}
 \varepsilon_1\ =\ \frac{1}{8}\left\langle \rho^2\right\rangle\ 
\end{equation}
is called the diamagnetic shift. For $S$-states, the experimental  measurement of $\varepsilon_1$   reveals physical properties such as the exciton mass, size, and spin \cite{Stier}. 
Taking our compact approximations, we calculated the diamagnetic shifts of the first $S$-states with $0\leq n_\rho\leq6$ with accuracy of  5  significant digits. Results are presented in Table \ref{groundstate}.
%
To calculate higher order corrections $\varepsilon_{n>1}$,  we used  the non-linearization procedure as described in \cite{Radial}. There, it was applied to construct the strong coupling expansion of the cubic anharmonic oscillator, see Subsection A.  In this way, 
we calculated higher corrections $\varepsilon_{n>1}$. Using as zero-order approximation (\ref{trial}), we established  corrections with  9 exact decimal digits. 
In Table \ref{TablePerturbation}, we present the first eleven  corrections in (\ref{Perturbation}) for the ground state. Following Dyson's argument, it is expected to be a divergent series.  However, numerical results suggest that the second term in (\ref{Perturbation}), namely
\begin{equation}
	\Delta \varepsilon(\gamma)\ =\  \sum_{n=1}^{\infty}\varepsilon_n\gamma^{2n}
\end{equation}
 is an alternating series and Pad\'e resummable. For example, using a Pad\'e approximant $P_{5}^6(\gamma^2)$  based on the first 11 corrections  leads to accurate results in the domain $0\leq\gamma\leq 1$ with 4 exact significant digits.  In Fig. \ref{Fig:MagnetoExcitons}, we presented the plots of $\Delta\varepsilon(\gamma)$ for the first seven $S$-states calculated through Pad\'e approximants $P_{5}^{6}(\gamma^2)$.

\begin{table}[h]
\caption{ Ground state (0,0): Numerical coefficients $\varepsilon_n$, $n = 0, 1, ..., 11$ of the series expansion (\ref{Perturbation}) for $\varepsilon$ calculated in non-linearization procedure. }
{\setlength{\tabcolsep}{0.35cm}
\begin{tabular}{|cccc|}
					\hline
					\hline
\rule{-1pt}{03ex}     
$n$ & $\varepsilon_n$ & $n$  & $\varepsilon_n$ \\[5pt]
	\hline
\rule{0pt}{4ex}0 &0.179\,935\,400            & 6  &$-0.018\,000\,868$             \\[5pt]
1 &0.136\,337\,679          & 7  &0.030\,034\,032         \\[5pt]
2 &$-0.023\,813\,717$            & 8  &$-0.057\,913\,120$            \\[5pt]
3 &0.0129\,299\,044             & 9  &0.127\,105\,525        \\[5pt]
4 &$-0.011\,145\,815$             & 10 &$-0.314\,135\,328$            \\[5pt]
5 &0.012\,765\,279             & 11 &0.867\,136\,787     \\[5pt]       
\hline
\hline
\end{tabular}}
\label{TablePerturbation}
\end{table}

	\section{Three-Particle System: Trions}
 \label{SectionII}
\begin{figure}[h]
		\includegraphics[width=0.3\textwidth]{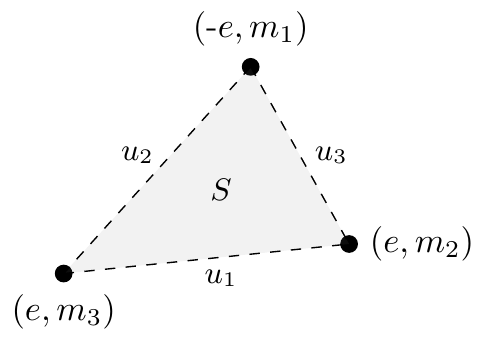}
		\caption{Two-dimensional complex of three particles. \textcolor{black}{Each  constituent particle is labeled by $(q,m)$, where $q$ is the charge while $m$ the mass.} The area enclosed, denoted by $S$ and shaded in gray,  is given by (\ref{Heron}). }
		\label{Helium}
	\end{figure}

 In this Section, we consider the lowest bound state of the two-dimensional complex  made of three charged particles of (effective) masses $(m_1, m_2, m_3)$ and charges $(-e, e, e)$, respectively.  The pairwise interaction between them is assumed to be logarithmic, and it is given by (\ref{potential}).  In Fig. \ref{Helium}, the geometrical setting of the system is shown.

 The ground  state  is a square-integrable eigenfunction with dependence only on the relative distances $u_1=r_{23}$, $u_2=r_{13}$, and  $u_3=r_{12}$, and it is the lowest eigenfunction of the Hamiltonian\footnote{The ranges
of the three variables $u_i$ are coupled and satisfy a \textit{triangle condition}: their lengths must be such that they can form a triangle \cite{Escobar}.} 
	\begin{equation}
		\hat{H}_{T}\psi\ =\ E\psi\ ,\qquad \psi\ =\ \psi(u_1,u_{2},u_{3})\ ,
  \label{SchrodingerTrion}
	\end{equation}
	\begin{align}
		\hat{H}_{T} 	=\ -\frac{\hbar^2}{2}\Delta\ -\ \frac{e^2}{\rho_0}\ln\left(\frac{u_1}{\rho_0}\right) \ +\ \frac{e^2}{\rho_0}\ln\left(\frac{u_2}{\rho_0}\right)+\ \frac{e^2}{\rho_0}\ln\left(\frac{u_3}{\rho_0}\right)\  
	\end{align} 
	with \cite{Turbiner3body}
	\begin{align}
		\Delta\ =&\ 
		\frac{1}{\mu_{23}}\frac{1}{u_{1}}\pa_{u_{1}}\left(u_{1}\pa_{u_{1}}\right)
		\ +\ 	\frac{1}{\mu_{13}}\frac{1}{u_{2}}\pa_{u_{2}}\left(u_{2}\pa_{u_{2}}\right)
		\ +\  \frac{1}{\mu_{12}}\frac{1}{u_{3}}\pa_{u_{3}}\left(u_{3}\pa_{u_{3}}\right)
		\nonumber \\
		&\  
		+\frac{1}{m_1}\frac{\ u_{2}^2+u_{3}^2-u_{1}^2}{u_{2} u_{3}}\pa_{u_{3}}\pa_{u_{2}}\
		+\frac{1}{m_2}\frac{u_1^2+\ u_{3}^2-u_{2}^2}{\,u_{1} u_{3}}\pa_{u_{1}}\pa_{u_{3}}
		\  +\frac{1}{m_3}\frac{\ u_{1}^2+u_{2}^2-u_{3}^2}{u_{1} u_{2}}\pa_{u_{1}}\pa_{u_{2}}\ .
		\label{relative}
	\end{align}
	where
	\begin{equation}
		\mu_{ij}\ =\ \frac{m_im_j}{m_i+m_j}\ , \qquad i,j=1,2,3\ .
  \label{r_mass}
	\end{equation}
 are the reduced masses. The operator $\hat{H}_T$ is self-adjoint with respect to the volume element 
	\begin{gather}	
		dV\ \propto\  u_1\,u_2\,u_3\,S^{-1}\,du_1\,du_2\,du_3\ , \nonumber \\ S\ =\  \frac{1}{4}\sqrt{(u_1+u_2+u_3)(u_1+u_2-u_3)(u_2+u_3-u_1)(u_1+u_3-u_2)}\ . 
  \label{Heron}
	\end{gather}    

We set $m_2=m_3$, as it \textcolor{black}{usually appears} in trions\footnote{\textcolor{black}{It has been established that for monolayers of TMDCs based of Tungsten, electrons in negative trions may have different effective masses, see \cite{Tuan}.}}.  
Under this assumption, 
	\begin{equation}
		\mu_{12}\  =  \mu_{13}\ ,\qquad \ \mu_{23}\ =\ \frac{m_{2,3}}{2}\ .
	\end{equation}
Next, we introduce the following change of variables
	\begin{equation}
		u_{i}\rar\left(\frac{\hbar^2\rho_0}{m_1\,e^2}\right)^{-\frac{1}{2}}u_i\ ,\qquad i=1,2,3\ ,
  \label{transformationTrion}
	\end{equation}
 and the parameter
  \begin{equation}
		\sigma\ =\ \frac{m_1}{m_{2,3}}
  \label{mass_ratio}
	\end{equation}
  to remove  in (\ref{SchrodingerTrion}) the appearance of physical constants (except for the masses $m_1$ and $m_{2}=m_3$). Therefore, the trion ground state wave function corresponds to the nodeless solution of the \textcolor{black}{dimensionless} equation
 \begin{equation}
 	-\frac{1}{2}\Delta_\sigma \psi\  +\ \ln\left(\frac{u_2u_3}{u_1}\right)\psi\ =\ \varepsilon(\sigma)\,\psi\ ,
 	  \label{trion_SE}
 \end{equation}
where 
	\begin{align}
\Delta_\sigma\ =\ 	&\frac{2\sigma}{u_{1}}\pa_{u_{1}}\left(u_{1}\pa_{u_{1}}\right)
		\ +\ 	\frac{\sigma+1}{u_{2}}\pa_{u_{2}}\left(u_{2}\pa_{u_{2}}\right)
		\ +\  \frac{\sigma+1}{u_{3}}\pa_{u_{3}}\left(u_{3}\pa_{u_{3}}\right)
		\nonumber
		\\
		&+\ \frac{\ u_{2}^2+u_{3}^2-u_{1}^2}{u_{2} u_{3}}\pa_{u_{3}}\pa_{u_{2}}\
		+\ \sigma\frac{u_1^2+\ u_{3}^2-u_{2}^2}{\,u_{1} u_{3}}\pa_{u_{1}}\pa_{u_{3}}
		+\  \sigma\frac{u_{1}^2+\ u_{2}^2-u_{3}^2}{u_{1} u_{2}}\pa_{u_{1}}\pa_{u_{2}}\ .
	\end{align}
	In (\ref{trion_SE}), $\varepsilon(\sigma)$ is related to the total energy ($E$), as follows
	\begin{equation}
		E\ =\ \frac{e^2}{\rho_0}\varepsilon(\sigma)\ -\ \frac{e^2}{2\rho_0}\ln\left(\frac{m_1e^2\rho_0}{\hbar^2}\right)\ .
  \label{energyT}
	\end{equation}
 In contrast to the energy levels of an exciton, the  difference in energy of two arbitrary levels does depend on the masses through  $\sigma$.

 \subsection{Trial Function}
Based on the functions (\ref{trial}) constructed for excitons in Section \ref{SectionI}, we design  compact trial functions for a trion in its ground state. 
	Since we assumed that \textcolor{black}{$m_2=m_3$, the  positively charged trion $(e^-,h,h)$ and the negative  one ($h,e^-,e^-$) are described by the same Schr\"odinger equation\footnote{\textcolor{black}{When the constituent particles have different masses, they are distinguishable. Consequently, no symmetry requirement is imposed to the total wave function. In particular, in (\ref{Kinoshita}) the symmetrizer is no longer needed.}}.   Their total wave functions, which include spin and valley quantum numbers, are antisymmetric with respect to the permutation of  identical particles, see \cite{Courtade}.  In particular,  the spatial part of the  ground state wave function    must be symmetric\footnote{According to our variational calculations, the anti-symmetric $S$-state is \textit{repulsive} and does not correspond to a bound state. This feature was already established for the Rytova-Keldysh potential in the non-logarithmic regime, see \cite{Courtade,Fey}. }  under this permutation. Therefore,} we propose the following symmetric wave function
	\begin{equation}
		\psi^{(approx)}(u_1,u_2,u_3)\ =\ \hat{S}\left[(1\ +\ \gamma\, u_1^2\ln(u_1^2))\,e^{-\Phi_{0,0}(\alpha^2\,u_{2})-\Phi_{0,0}(\beta^2\,u_{3})}\right] \ ,
  \label{Kinoshita}
	\end{equation} 
\textcolor{black}{ where the symmetrizer operator is given by
 \begin{equation}
   \hat{S}\ =\ \frac{1}{2}(1\ +\ \hat{P}_{23})\ .  
 \end{equation}
Here $\hat{P}_{23}$ is the permutation between the identical particles $2\rar3$ in $u$-variables.
 The form of the {\it phase} is given in (\ref{phase}).}
 Note that $\Phi_{0,0}$ is given by formula (\ref{phase}) with optimized  parameters found in Table \ref{groundstate}. The construction of the previous wave function is motivated as follows. The factor $e^{-\Phi_{0,0}(\alpha^2\,u_{2})-\Phi_{0,0}(\beta^2\,u_{3})}$ describes the trion when the repulsive term of the interaction is absent. In turn, the factor $ 1+\gamma u_{1}^2\ln(u_{1}^2)$ takes into account the polarization of the complex and describes the repulsion between equally charged carriers. \textcolor{black}{This fact can be easily established using asymptotic analysis in (\ref{trion_SE}) at small $u_{1}$, which leads to the expansion
\begin{equation}
  \psi(u_1,u_2,u_3)= 1\ -\frac{1}{4\sigma}\ u_1^2\ln (u_1)+\ \frac{1}{2 (\sigma +1)}(u_2^2\ln (u_2)+\ u_1^2\ln (u_3))\ +\ ...
 \ .
\end{equation} }
 Parameters $\{\alpha,\beta\}$ were introduced in such a way that (i) they admit the meaning of screening charges if $m_1$ is kept fixed; and (ii) the  square-integrability of the trial function is guaranteed.
 	 Altogether, the trial function (\ref{Kinoshita})  contains only three free real parameters: $\{\alpha,\beta,\gamma\}$. 
	   To fix their values, we use the variational method to find the optimal configuration that minimizes the variational energy. Numerical calculations were performed in perimetric coordinates using a  modification of  the FORTRAN code  described in \cite{TurbinerCompact}. 
	   
	   We carried out calculations for $\sigma\in[0,10]$, thus, covering  representative TMDCs (see below). In Fig. \ref{Optimal}, we present the optimal parameters as functions of $\sigma$ in the domain $\sigma\in[0,10]$. They have a smooth behavior, especially at large $\sigma$. We found that $\beta>\alpha$ in the  domain considered. It hints that one of the two equally charged carriers  is closer to the opposite-charged one. Compared to $\alpha$ and $\beta$, parameter $\gamma$ is  smaller. It reaches it maximum value at $\sigma\sim1-2$.  In Table \ref{energies_trion}, optimized variational energy $\varepsilon(\sigma)$ is presented. 
\begin{figure}[h]
	\centering
	\begin{minipage}{0.45\textwidth}
		\centering
		\includegraphics[width=1\textwidth]{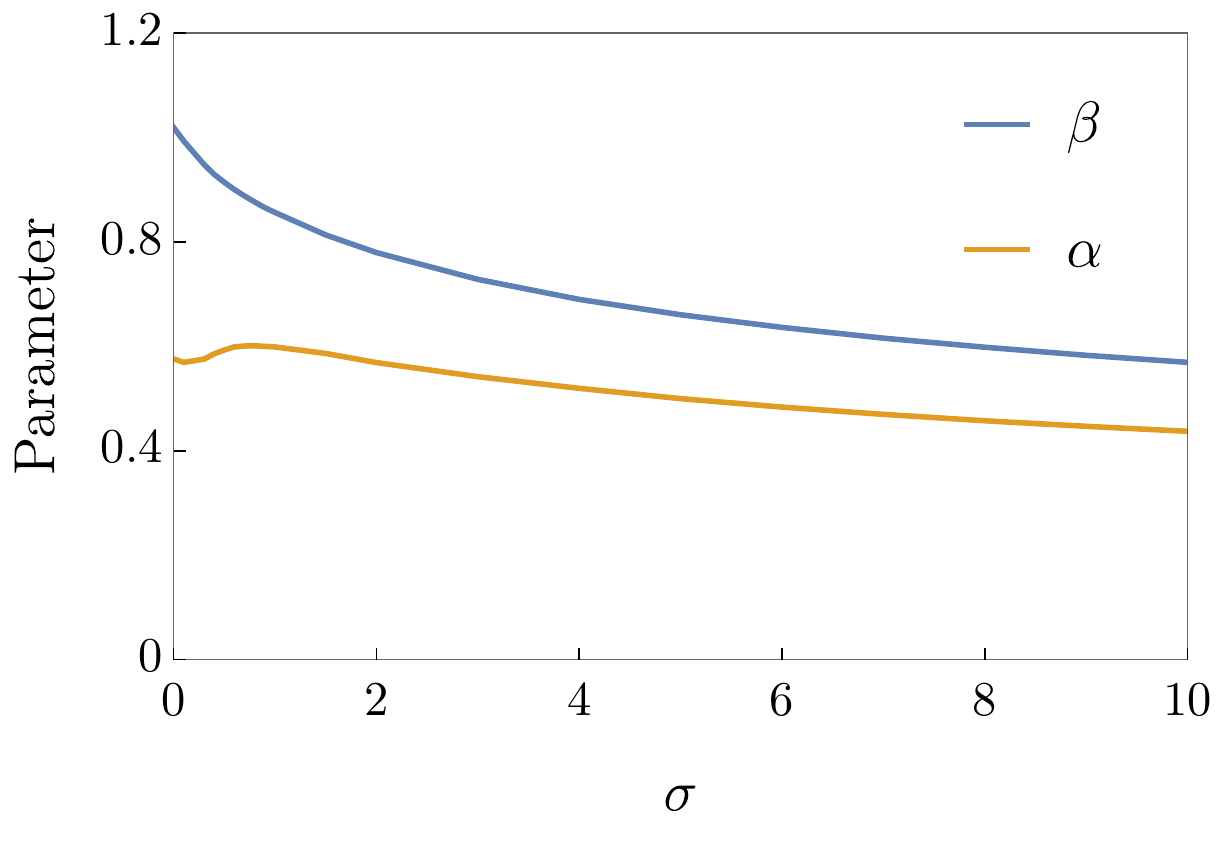} 
	\end{minipage}\hfill
	\begin{minipage}{0.45\textwidth}
		\centering
		\includegraphics[width=1\textwidth]{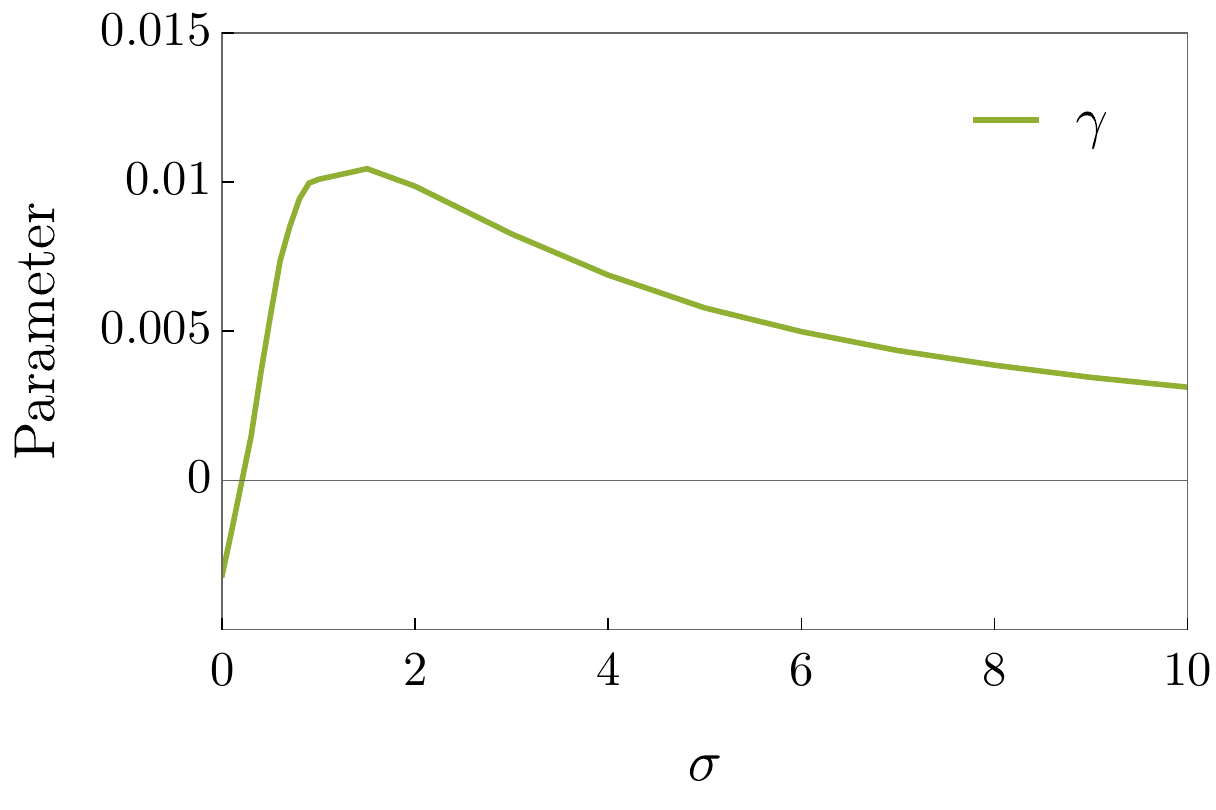} 
	\end{minipage}
		\caption{\textcolor{black}{Dimensionless} optimal variational parameters of  Ansatz (\ref{Kinoshita}) as functions of  the mass ratio $\sigma$. }
		\label{Optimal}
\end{figure}

\textcolor{black}{
\begin{table}[h]
		\caption{Trion dimensionless energy $\varepsilon(\sigma)$ vs the mass ratio $\sigma$, see (\ref{energyT}). Variational energies obtained through trial functions (\ref{Kinoshita}) and (\ref{improved}). For the latter, calculations correspond to  $N=1$, which were confirmed with $N=2$. Results rounded to first display decimal digits.}	
			{\setlength{\tabcolsep}{0.5cm}
{\color{black}\begin{tabular}{|ccc|ccc|}
\hline\hline
&\multicolumn{2}{c|}{\rule{0pt}{3ex} Wave function}&    & \multicolumn{2}{c|}{\rule{0pt}{3ex} Wave function} \\[4pt]
\cline{5-6}
\cline{2-3}
$\sigma$ & Eq. (\ref{Kinoshita})  & Eq. (\ref{improved}) & $\sigma$& \rule{0pt}{3ex}  Eq. (\ref{Kinoshita})  &  Eq. (\ref{improved})  \\[5pt]
 \hline
 \rule{0pt}{4ex}0   & 0.0871  &0.0791               & 1   &0.42265     &0.4182              \\[5pt]
0.1    &0.1365  &0.1285             & 2    &0.6091    &0.6058               \\[5pt]
0.2    &0.1809  &0.1729             & 3    &0.7450    &0.7411                \\[5pt]
0.3    &0.2202  &0.2165             & 4    &0.8506    &0.8473               \\[5pt]
0.4    &0.2563  &0.2539             & 5    &0.9378    &0.9347               \\[5pt]
0.5    &0.2891  &0.2868             & 6    &1.0117    &1.0090                \\[5pt]
0.6    &0.3194  &0.3166             & 7    &1.0763    &1.0737               \\[5pt]
0.7    &0.3476  &0.3443             & 8    &1.1333    &1.1308               \\[5pt]
0.8    &0.3740  &0.3703             & 9    &1.1845    &1.1821               \\[5pt]
0.9    &0.3990  &0.3949             & 10   &1.2309    &1.2286           \\[5pt]   
\hline\hline
\end{tabular}}}
\label{energies_trion}
\end{table}}


\textcolor{black}{\subsection{Alternative Trial Function}
To investigate the accuracy of the variational calculations shown in the previous Section, we use an alternative trial function to estimate the energy of the trion in the lowest $S$-state with higher accuracy. In a similar way it was done for the exciton, see (\ref{hydrogen}), we proposed
\begin{equation}
 \psi(u_1,u_2,u_3)= \hat{S}\left[\,\sum_{i,j,k=0}^N c_{i,j,k}\, u_{1}^{2i}u_{2}^{2j}u_{3}^{2k} (1\ +\ \gamma\, u_1^2\ln(u_1^2))\,e^{-\Phi_{0,0}(\alpha^2\,u_{2})-\Phi_{0,0}(\beta^2\,u_{3})}\right]\ ,
 \label{improved}
\end{equation}
 where $c_{i,j,k}$ are coefficients to be determined. Without loss of generality, we set $c_{0,0,0}=1$ as normalization of the approximate wave function.  Note that at $N=0$, trial functions (\ref{Kinoshita}) and (\ref{improved}) coincide. Meanwhile at $N>0$,    they differ due to  the insertion of the polynomial prefactor $\sim u_1^{2i}u_2^{2j}u_3^{2k}$. This insertion corrects the behavior of the wave function at small distances, leading to a higher accuracy for the ground state energy. To fix the value of the parameters $\{\alpha,\beta,\gamma\}$ and $c_{i,j,k}$'s, we followed a two-step variational consideration. First, we fix the value of parameters $\{\alpha,\beta,\gamma\}$ according to the variational calculations of the previous Section, in which the polynomial prefactor was not considered. The remaining free parameters, $c_{i,j,k}$'s, are defined by means of the secular equations. To solve them, we used
the L\"owdin orthogonalization procedure. In Table \ref{energies_trion}, we present the value of $\varepsilon(\sigma)$ calculated with $N=1$ in (\ref{improved}) for representative values of $\sigma$. Using $N=2$, we confirmed the results for $N=1$ up to the first four decimal digits. It indicates a fast rate of convergence with respect to $N$. In this way, we concluded that the wave function (\ref{Kinoshita}) provides an accuracy of two decimal digits for all $\sigma>0$.
\\ \\
The plot of $\varepsilon(\sigma)$, calculated through trial function (\ref{improved}) with $N=1$, is shown in Fig. \ref{Helium_energy}. The curve described by $\varepsilon(\sigma)$ can be easily interpolated with an accuracy of 2  decimal digits by the simple function
\begin{gather}
\varepsilon_{fit}(\sigma)\ =\ a\  + \ b\,\ln(1+c\,\sigma)\ ,\nonumber\\
a\ =\ 0.083\ ,\qquad b\ =\ 0.465\ ,\qquad c\ =\ 1.059\ , 
\label{fit}
\end{gather}
in the domain $\sigma\in[0,10]$.  
	\begin{figure}[h]
		\includegraphics[width=0.55\textwidth]{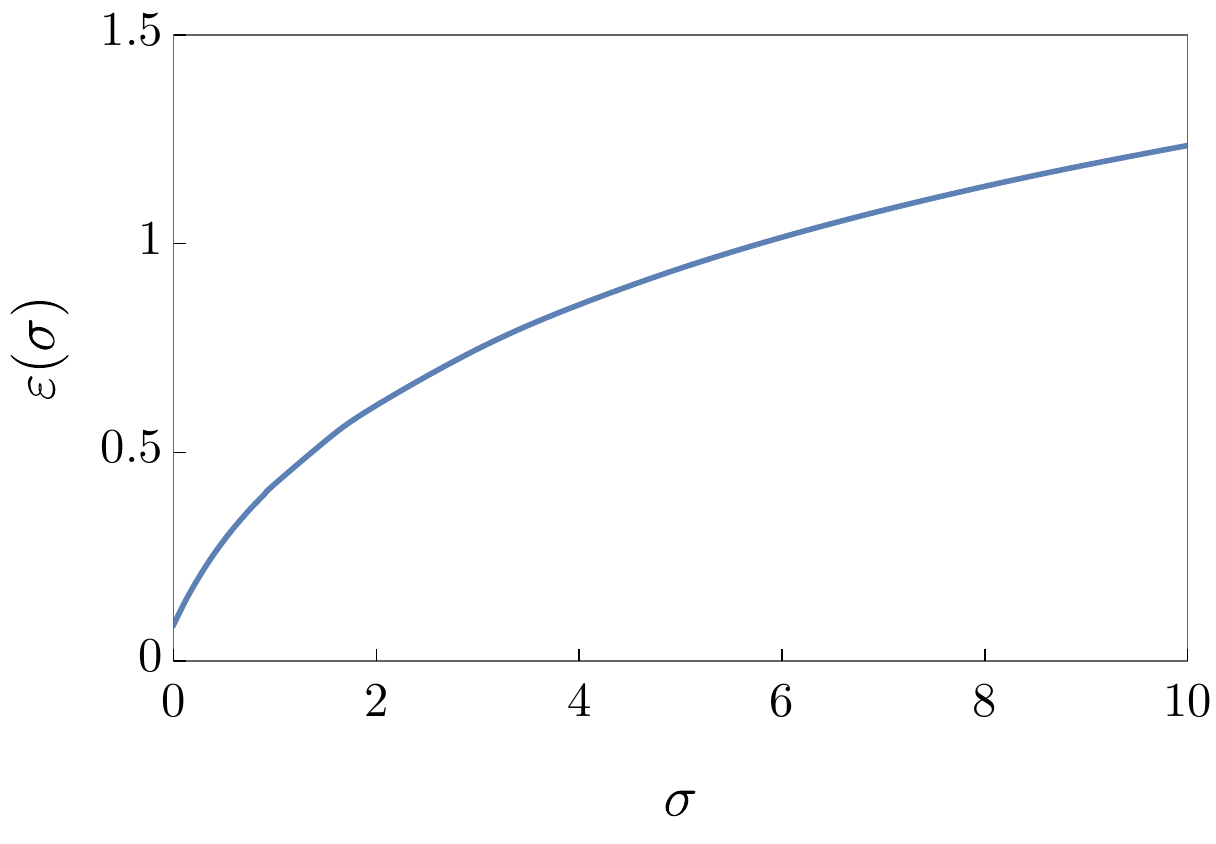}
		\caption{Plot of \textcolor{black}{dimensionless energy of trions $\varepsilon$}, see (\ref{energyT}),  as a function of the mass ratio $\sigma$.  }
		\label{Helium_energy}
	\end{figure}
 }

 \textcolor{black}{\subsection{Comparison with Experimental Data}}
 In order to compare our results with experimental data, we calculated the binding energy ($E_b$).  It is defined as the amount of energy needed to dissociate a trion into a neutral complex (exciton) and a free carrier. Therefore, 
\begin{equation}
 E_{b}\ =\     |E_{trion}-E_{exciton}|\ ,
\end{equation}
where the energies correspond to ground states.
According to (\ref{energyE}), (\ref{energyT}), and (\ref{fit}), we have a compact expression for $E_b$, 
\begin{equation}
 E_{b}(\sigma)\ =\   \frac{e^2}{\rho_0}\left( 0.179935-\varepsilon_{fit}(\sigma)+\frac{1}{2}\ln(1+\sigma)\right)\ .
 \label{bindingfit}
\end{equation}
Hence, the binding energy for trions is a function of $\sigma$. 


In Table \ref{energies_trion_exp}, we compare  theoretical binding energies given by (\ref{bindingfit}) with experimental ones for negatively charged trions\footnote{Thus, we take $m_1=m_h$ and $m_2=m_e$.} in molybdenum (Mo) and tungsten (W) dichalcogenide materials. As previously established \cite{Mayers}, 
we found that  logarithmic potential overbinds the trion, leading to larger binding energies than those reported in experiments.
However, our results do not confirm   the statement that the logarithmic potential leads to binding  energies  50\% larger, see \cite{Mayers}.
The largest deviation occurs for MoSe$_2$, reaching a deviation in energies of 30\%. In turn, the smallest deviation is found for WS$_2$, for which is 16\%. Interestingly, either formula (\ref{bindingfit}) or
experimental results predicts equal binding energy for MoS$_2$ and WS$_2$.

  \begin{table}[h]
		\caption{Negative charged trion binding energy in meV  for different suspended TMDC monolayers. \textcolor{black}{Results rounded to first displayed significant digits.} Values of $\rho_0$, $\sigma$, and experimental binding energies   taken from reference \cite{Szyniszewski}. 	}
			{\setlength{\tabcolsep}{0.30cm}
\begin{tabular}{|c|cccc|}
\hline\hline
\multirow{2}{*}{\rule{0pt}{2ex}Material} &             &    & \multicolumn{2}{c|}{\rule{0pt}{3ex}$E_b$ (meV)} \\
\cline{4-5}
 & $\rho_0$ (\AA) & $\sigma$& \rule{0pt}{3ex}Present Results    & Experiments    \\[5pt]
\hline
  \rule{0pt}{4ex}MoS$_2$      &39          &0.81 &41&34, 35   \\[5pt]        
   MoSe$_2$    &40          &0.86&39 &30     \\[5pt]             
    WS$_2$  &38          &0.84&41 &34, 36        \\[5pt]          
    WSe$_2$  &45          &0.85 &35&30     \\[5pt]                            \hline\hline
\end{tabular}}
\label{energies_trion_exp}
\end{table}

	\section{Toward the Rytova-Keldysh potential}
	\label{SectionIII}
If  $\rho_0$ in (\ref{potential}) is comparable to the  exciton Bohr radius, the logarithmic potential (\ref{potential}) is no longer suitable to describe the interaction between carriers inside monolayers of TMDCs. Furthermore,  it overbinds three-particle complexes as we have seen.  Under these circumstances, an adequate description of the interaction is given by the celebrated  Rytova-Keldysh potential \cite{Rytova,Keldysh}, namely 
\begin{equation}
	V_{RK}(\rho)\ =\ \frac{\pi q_iq_j}{2\rho_0}W\left(\frac{\rho}{\rho_0}\right)\  ,
	\label{eq:potential_interaction}
\end{equation}
where 
\begin{equation}
	W\left(\frac{\rho}{\rho_0}\right)\ =\ H_0\left(\frac{\rho}{\rho_0}\right)\ -\  Y_0\left(\frac{\rho}{\rho_0}\right)\ .
\end{equation}	
Here $H_0$ and $Y_0$ are the Struve and  Bessel function of second kind, respectively.
\textcolor{black}{At small $\rho$, the Rytova-Keldysh potential is reduced to the logarithmic potential as follows 
\begin{equation}
V_{RK}(\rho)\ =\ -\frac{q_iq_j}{\rho_0}\left(\ln\left(\frac{\rho}{\rho_0}\right)\ +\ \Gamma \ -\ \frac{\rho}{\rho_0} +\mathcal{O}(\rho^2\ln\rho)\right)\ ,\qquad \rho\rar0\ ,
\end{equation}
where $\Gamma=\gamma-\ln2$; here $\gamma$ denotes the Euler-Mascheroni constant. The constant $\Gamma$ fixes the reference point for the energy. Therefore, it plays no relevant physical role when studying binding energies of complexes interacting within the logarithmic regime. On the other hand, at large distances 
\begin{equation}
V_{RK}(\rho)\ =\ \frac{q_iq_j}{\rho}\left(1\ -\ \left(\frac{\rho_0}{\rho}\right)^2\ +\ 9\left(\frac{\rho_0}{\rho}\right)^4\ +\ \mathcal{O}(\rho^{-6})\right)\ ,\qquad \rho\rar \infty\ . 
\end{equation}
Therefore, in this limit, the dominant interaction is given by the Coulomb one. 
The radial  Schr\"odinger equation that describes excitons now reads
\begin{equation}
-\frac{\hbar^2}{2\mu}\left(\pa_\rho^2\psi\ +\ \frac{1}{\rho}\pa_\rho\psi\right)\ +\ \left(\frac{\hbar^2m^2}{2\,\mu\rho^2}\ -\ \frac{\pi e^2}{2\rho_0}W\left(\frac{\rho}{\rho_0}\right)\right)\psi\ =\ E\,\psi\ .
\end{equation}
  To construct a compact trial function for excitons, we can follow the  approach presented in Section I.B, in which the main ingredient is the asymptotic series.
 Using the transformation (\ref{transformation}), we obtain the dimensionless Schr\"odinger equation 
\begin{equation}
-\frac{1}{2}\left(\pa_\rho^2\psi\ +\ \frac{1}{\rho}\pa_\rho\psi\right)\ +\ \left(\frac{m^2}{2\,\rho^2}\ -\ \frac{\pi\la}{2}W\left(\la\rho\right)\right)\psi\ =\ \varepsilon\,\psi \ .
\label{adimensional}
\end{equation}
where\footnote{Here $a_0$ denotes the exciton Bohr radius defined as $a_0=\dfrac{\hbar^2}{\mu\,e^2}$. Meanwhile, $\mathcal{E}_0$ represents the exciton Rydberg constant. } 
\begin{equation}
\la\ =\ \frac{a_0}{\rho_0}\ ,  \qquad\varepsilon\ =\ \frac{E}{\mathcal{E}_0}\ ,\qquad \mathcal{E}_0\ =\ \frac{\mu\,e^4}{\hbar^2}\ .
\label{coupling}
\end{equation}
Note that $\la$ corresponds to the ratio of the two length scales of the system. According to parameters found in \cite{Szyniszewski}, $\la\lesssim 8\times10^{-2}$ for all materials presented in Table \ref{energies_trion_exp}. Therefore, it suggests taking the potential  in equation (\ref{adimensional}) and expanding it in powers of $\la$,
\begin{equation}
-\frac{\pi\la}{2}W(\la\rho)\ =\ \la\left(\,\ln(\la\rho)\ +\ \Gamma\ -\ \la\rho\ +\ \mathcal{O}(\la^2)\right) \ ,\qquad \la\rar0
\end{equation}
Therefore, the logarithmic potential is dominant\footnote{The same conclusion  for the $N$-particle complex can be established using similar arguments. } at small $\la$. 
For arbitrary $\la$,  an asymptotic analysis establishes that the phase of the exciton wave function has the asymptotic series 
 	\begin{equation}
 	\Phi_{n_\rho,m}(\rho)\ =\   \sum_{i=1}^{\infty}\left\{\rho^{2i}\sum_{j=0}^ia_{ij}\ln^{j}(\rho)\ +\ \rho^{2i+1}\sum_{j=0}^{i-1}b_{ij}\ln^{j}(\rho)\right\}\ ,\qquad \rho\ \rar\ 0\ ,
 \end{equation}
where $a_{i,j}$ and $b_{ij}$ are coefficients, cf. (\ref{series1}).	
Consequently, the wave function behaves as
\begin{equation}
\psi_{n_{\rho,m}}(\rho)\ =\ \rho^{|m|}\left(1\ +\ \sum_{i=1}^\infty\left\{\rho^{2i}\sum_{j=0}^iA_{ij}\ln^{j}(\rho)\ +\ \rho^{2i+1}\sum_{j=0}^{i-1}B_{ij}\ln^{j}(\rho)\right\}\right)\ ,\qquad \rho\ \rar\ 0\ ,
\label{smallRK}
\end{equation}
with $A_{ij}$ and $B_{ij}$ coefficients.
At large distances, the phase acquires the following form
\begin{equation}
\Phi_{n_r,m}(\rho)\ =\ \sqrt{-2\varepsilon}\,\rho\ +\ \frac{1}{2} (4n_\rho+2m+1)\ln\rho \ +\ \sum_{k=1}^{\infty}c_k\rho^{-k}
\label{largeRK}
\end{equation}
where $c_k$'s are coefficients. In contrast with (\ref{series3}), for the Keldysh-Rytova it is possible to find the structure exciton wave function at large distances.}
 To the best of the authors' knowledge,  this piece of information is absent in the literature, and therefore it has not been  exploited for the construction of trial functions, see for example \cite{Semina}. Certainly, it will lead to an improvement in terms of local accuracy of wave function and variational energies compared with the current trial functions based on hydrogen-like orbitals, see \cite{Martins}.  
 \textcolor{black}{In our present approach, the next step to construct a compact wave function is the interpolation between series (\ref{smallRK}) and (\ref{largeRK}). A minimal interpolation results in
 \begin{equation}
 \Phi_{n_\rho,m}^{(approx)}(\rho)\ =\ \frac{A\ +\ B\rho^2\ln(1+C\rho)\ +\ D\rho^3}{\sqrt{1\ + \ F\rho^2\ln\rho\ +\ G\rho^4}}   \ +\ \frac{1}{8}(4n_\rho+2m+1)\ln(1\ +\ G\rho^4)	\ .
\end{equation}
where $\{A,B,C,D,F,G \}$ are free parameters. Once introduced in (\ref{representation}), it leads to the approximate form of the exciton wave function. Its accuracy in the framework of the variational method will be studied elsewhere.}
 Finally, the three-particle  compact wave functions can be constructed following the same procedure shown in this work: going from two to three particles using as building block the exciton wave function.
 
 \section{Conclusions}
 In this article,  locally accurate wave functions were constructed for  the bound states of two-particle system (exciton) in two-dimensions whose constituent particles interact through a logarithmic potential.  For states with quantum numbers $n_{\rho}\leq4$ and $|m|\leq5$, they allowed us to reproduce energies with 5-6 exact decimal digits. It was checked using the non-linearization procedure and an alternative two-step variational approach.   For magnetoexcitons at rest, it was demonstrated that those functions (used as a zero-order approximation) lead to highly accurate coefficients of  the weak coupling energy expansion in powers of the magnetic field strength. Numerical results for $S$-states, suggest that the weak coupling expansion is Pad\'e (re)summable.   
 
 We constructed a compact approximation of the two-dimensional three-particle system (trion) ground state wave function  using as a building block the exciton ground state one.
 \textcolor{black}{It only depends  on three non-linear free parameters that are fixed through the variational method. It contrasts with the $10^{6}$ linear variational parameters that were recently used in  \cite{Fey}.}
 Our wave function led to binding energies in good  agreement  with experimental results for trions in concrete   TMDCs made of Mo, W, S, and Se. It was shown that the logarithmic potential leads to binding energies $\lesssim30\%$  larger than those coming from experiments. We find a simple formula for the binding energy as a function of the mass ratio of the constituent particles.

Finally, the structure of the exciton wave function at small distances whose carriers interact via the Rytova-Keldysh potential was established. This new piece of information may lead to the construction of improved  variational wave functions used to study larger complexes in TMDCs  \textcolor{black}{like bi-excitons}. 

	\section*{Acknowledgments}
	
	The authors thank Prof. A.V. Turbiner for drawing our attention to logarithmic interactions as well as valuable comments and suggestions. We thank J.C. L\'opez-Vieyra for the support with the variational calculations. J.C. del Valle thanks D.A Bonilla-Moreno for useful remarks and discussions. \textcolor{black}{The authors thank M. Szyniszewski for his interest and for providing additional information on Ref. \cite{Szyniszewski}. During the last stage of this work,  J.C. del Valle was partially supported  by the SONATABIS-10 grant no. 2019/34/E/ST1/00390. D.J.N acknowledges support by Fulbright COMEXUS Project NO. P000003405. }

	\bibliography{biblio} 

\end{document}